# Effectiveness Of Defect Prevention In I.T. For Product Development

**Suma V[1], Dr. T. R. Gopalakrishnan Nair [2]**

[1] Asst. Professor, Department of Information Science and Engineering, Dayananda Sagar College of Engineering, Bangalore. email: sumavdsce@gmail.com
[2] Director, Research and Industry Incubation Centre, Dayananda Sagar Institutions, Bangalore. email: trgnair@ieee.org

Dr. T. R. Gopalakrishnan Nair
Director,
Research and Industry Incubation Centre,
Dayananda Sagar Institutions, Bangalore-
560078, India
email: trgnair@ieee.org

**ABSTRACT**

Defect Prevention is the most critical but most neglected component of the software quality assurance in any project. If applied at all stages of software development, it can reduce the time, cost and resources required to engineer a high quality product.

Software inspection has proved to be the most effective and efficient technique enabling defect detection and prevention. Inspections carried at all phases of software life cycle have proved to be most beneficial and value added to the attributes of the software.

Work is an analysis based on the data collected for three different projects from a leading product based company. The purpose of the paper is to show that 55% to 65% of total number of defects occurs at design phase. Position of this paper also emphasizes the importance of inspections at all phases of the product development life cycle in order to achieve the minimal post deployment defects.

3-8PAGES

# Effectiveness Of Defect Prevention In I.T. For Product Development

## Introduction

The IT industry is successful, if it can gain the total satisfaction of the customers in every transaction. This is achievable if the organization can produce a high quality product. To identify a product to be of high quality, it should be free of defects, should be capable of producing expected results. It should be delivered in an estimated cost, time and be maintainable with minimum interferences.

A small increase in prevention measure will normally create a major decrease in total quality cost. But the main objective of quality cost analysis is not to reduce the cost, but to make sure that the cost spent are the right kind of cost and that maximize the benefit derived from that investment. Due to quality cost analysis, the major emphasis has been shifted to prevention of defects. Also it is observed in all IT companies over a period of time, at some optimum point, the business performance enhances, quality increases and the cost of quality decreases due to the adoption of defect detection and prevention activities.

A defect refers to any blemish or imperfection in a software work product or software process [1]. The term defect refers to an error, fault or failure. The IEEE/Standard defines the following terms as Error: human actions that lead to incorrect result. Fault: incorrect decision taken while understanding the given information, to solve problems or in implementation of process. Failure: is inability of a function to meet the expected requirements [2] [3].

## Need for defect prevention

Analysis of the defects at early stages reduces the time, cost and the resources required. The knowledge of defect injecting methods and processes enable the defect prevention. Once this knowledge is used appropriately, the quality is improved. It also enhances the total productivity.

## Benefits of defect prevention

There is need in all IT companies to reduce as many numbers of defects as possible. Existences of defect prevention strategies reflect a high level of test process maturity [6]. Detection of errors in the development life cycle helps to prevent the migration of errors from requirement specifications to design and from design into code. Defect prevention provides the greatest cost and schedule savings over the duration of the application development efforts. Thus, it significantly reduces the number of defects, brings down the cost for rework, makes it easier to maintain port and reuse, makes the system reliable, and offers reduced time and resources required for the organization to develop high quality systems. Defects can be traced back to the life cycle stages in which they were injected based on which the preventive measures are identified and also increase productivity. A preventive measure for defects is a mechanism for propagating the knowledge of lessons learned between projects.

### Approaches to identify defects

There are several approaches to identify the defects like inspections, prototypes, testing and correctness proofs [4].Formal inspection is a quality assurance technique for identifying defects at the early stages of the development. Through prototyping several requirements are clearly understood which helps in overcoming the defects. Testing is one of the least effective techniques. Those of the defects, which could have escaped from identification at the early stages, can be detected at the time of testing. Correctness proofs are also a good means of detecting defects especially at the coding stage. Achieving the correct construction is the most effective and economical method of building the software.

Among several approaches, inspection has proven to be the most valuable and competent technique in defect detection and prevention [4][13][14][15]. Inspection examines all software related artefacts to detect and eliminate defects before they get manifested. It also verifies the standard of excellence for software engineering artefacts [7][8].Inspection is carried out even before the implementation while testing is carried out after the realization of the artefacts. Inspection is a static technique of fault detection and removal which certainly reduces the number of defects [3].

### Inspection Metrics

Metrics are measured numerical values used to quantify the process and the product [11]. They are used to monitor the effectiveness of the process and the quality of the product. They also serve as a criteria based upon which the inspection planning can be improved [12]. The most required metric for the estimation of defects in the software development are

Inspection Yield = Total Defects Found / Estimated Total Defects * 100 %

Defect Removal Efficiency (DRE) is a measure for the defect removal ability in the development process, which can be calculated for the entire life cycle or at each phase of the development life cycle. When DRE is used at front end (before the code integration), it is called as early defect removal and when used at specific phase, it is referred to as phase effectiveness [5][16].Thus DRE is also computed as

DRE (%) = (Defects removed during development phase) *100/defects latent in the product

Latent defects are the sum of defects removed during the phase and the defects found late.

## CASE STUDY

A study was made in a leading product based company on projects of various capabilities and subsequent CASE study shows a sample of such three projects which is a practical work carried out and not a theoretical simulation.

**Analysis of inspection process in various life cycle phases of product development**

Inspection is carried out at every phase of the software development in order to uncover the maximum number of defects.

During the requirements phase, the product manager understands analyses and validates the product requirements. The artefact undergoes inspection by a person from quality office who is well experienced with analysis and testing along with the development team.

During the design phase, the high-level design and low-level design are the artefacts to be inspected.

Implementation phase begins with the write up of test cases. Inspection is carried out on test cases at the first place to detect the defects so that the source code generation will be free of such imperfections. Next phase inspects the generated code through peer review process.

**Observation**

Table.1. depicts the data analyzed for three different projects sampled out from a leading product based company. It depicts the estimated and actual product development time and defect summary.

Table.1. Analysis of three projects from a leading product based company

| For complete project (In Man hours) | Project1 | | Project2 | | Project3 | |
|---|---|---|---|---|---|---|
| | Estimated | Acutal | Estimated | Acutal | Estimated | Acutal |
| Total Project Time (*) | 340 | 370 | 802 | 826 | 540 | 580 |
| Total Requirments Time | 60 | 70 | 154 | 179 | 93 | 108 |
| Total Number of Defects | 20 | 30 | 51 | 77 | 31 | 46 |
| Total Design Time | 120 | 140 | 282 | 329 | 192 | 224 |
| Total Number of Defects | 40 | 65 | 94 | 153 | 64 | 104 |
| Total Implementation Time | 60 | 50 | 154 | 128 | 93 | 77 |
| Total Number of Defects | 10 | 11 | 26 | 28 | 15 | 17 |
| Total Inspection Time | 40 | 44 | 99 | 108 | 62 | 69 |
| Total Test Time | 40 | 58 | 99 | 140 | 62 | 91 |

(*) Total project time includes documentation, training, release time and other such parameters which are not of interest.

Table. 2. depicts the probabilities of defects at each phase of product development life cycle. From the Table .2. , it can be proven that on an average 25% to 30% of total amount of defects are observed at requirements phase of the development life cycle. Nearly 55% to 65% of defects are observed at design phase and 10% to 13% are seen at implementation phase of the development. This accentuates the necessity of inspection to be carried out at all phases of product development and more emphasize to be given at design phase. This technique prevents the defects from getting propagated and manifested if left uncovered.

Table.2. Probability of Defects at all phases of product development life cycle.

| Probabality of Defect | Project 1 | Project 2 | Project 3 |
|---|---|---|---|
| % of Defects at Requirments Phase | 28% | 30% | 28% |
| % of Defects at Design Phase | 61% | 59% | 62% |
| % of Defects at Implmentation Phase | 10% | 11% | 10% |

Defect prevention methodologies cannot always prevent all defects from entering into the applications because application is very complex and it is impossible to catch all the errors. Remaining defects resides in the product as residual or latent defects.

The Figure.1. is a graphical representation of percentage of defects observed in the various phases of product development life cycle

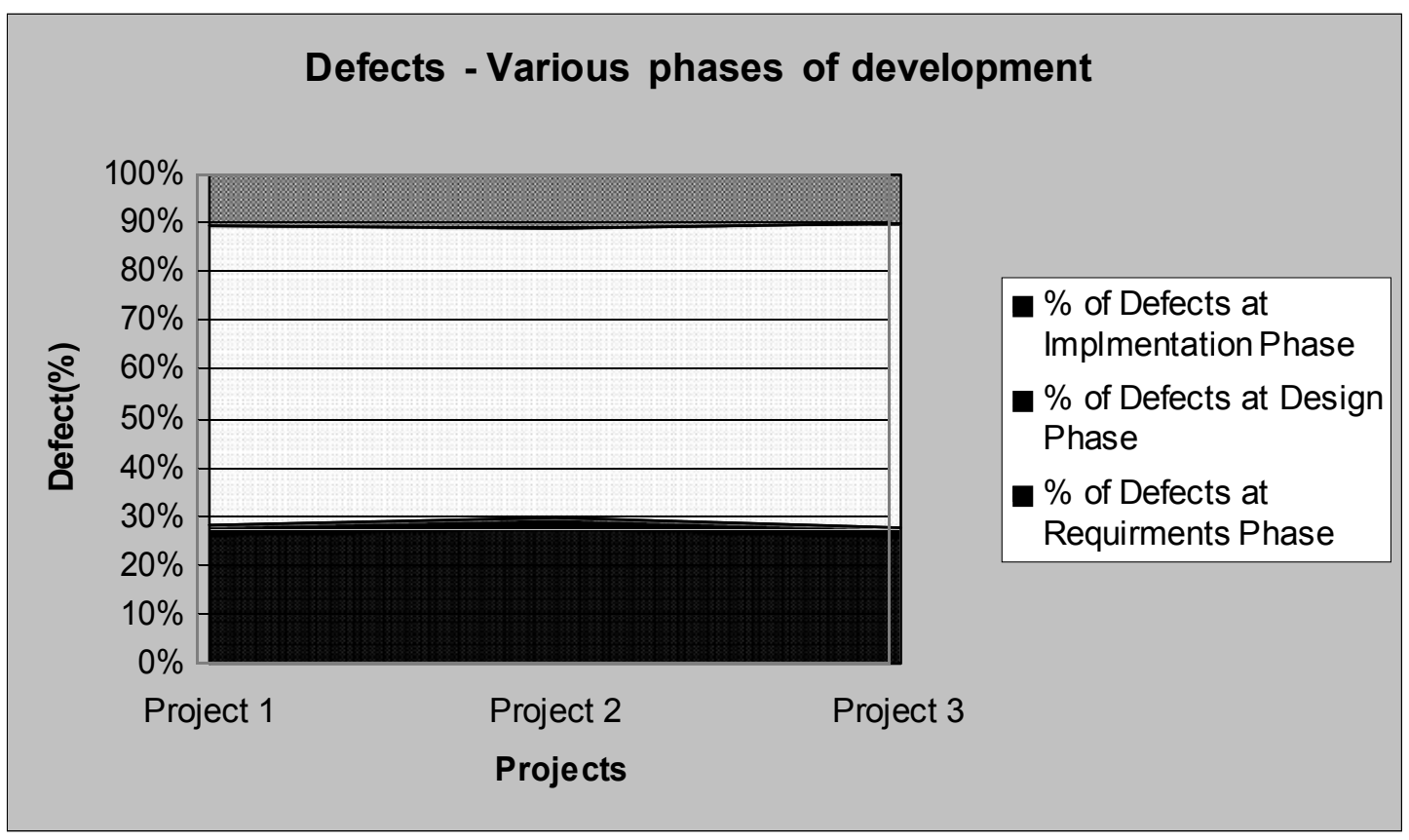

Figure. 1. Percentage of defects at each phase of product development cycle

## Benefits of Inspection

The benefits of software inspections are well identified since 1976.

- It is proven to be cost effective, as defects get uncovered before their migration to later phases of development
- It enhances the total productivity
- It increases customer satisfaction at all levels
- It adheres to meet the committed schedules
- It adds value to the dependency attributes of software like maintainability, availability and reliability [9]
- It reflects the maturity level of the company
- It can build up team spirit [10]
- Since inspection uncovers all static defects, time required for testing is reduced.

## Conclusion

Defect prevention is very much vital for an organization's quality growth. The main objective of quality cost is not to reduce the cost but to invest the cost on right investment.

There are several methods, tools, techniques and practices for defect prevention. Inspection is one of the most effective and efficient technique followed in all matured companies for the detection of defects in the early phases of the development life cycle and it also prevents defects to propagate in to the future.

Due to the varied benefits of inspections, it is always advisable to carry them at every phase of life cycle of software engineering. It is observed that 55% to 65% of total defects are at design phase and hence with inspection most of the defects can be uncovered.

**References:**

[1] Brad Clark, Dave Zubrow, "How Good Is the Software: A review of Defect Prediction Techniques" version 1.0, pg 5, sponsored by the U.S. department of Defense 2001 by Carnegie Mellon University.
[2] The Software Defect Prevention /Isolation/Detection Model drawn from www.cs.umd.edu/~mvz/mswe609/book/chapter2.pdf
[3] Jeff Tian "Quality Assurance Alternatives and Techniques: A Defect-Based Survey and Analysis" SQP Vol 3, No 3/2001, ASQ by Department of Computer Science and Engineering, Southern Methodist University.
[4] S.Vasudevan, "Defect Prevention Techniques and Practices" proceedings from 5th annual International Software Testing Conference in India, 2005.
[5] Stephan H. Kan, "Metrics and Models in Software Quality Engineering", 2nd edition Addison- Wesley professional, chapter 4- in-process quality metrics, December 2002
[6] Elfriede Dustin, Jeff Rashka, John Paul "Automate Software Sting" Chapter 4 Automate testing Introduction Process pg 144,ISBN 7-89494-044-5.
[7] Don O'Neill "Peer Reviews Encyclopedia of Software Engineering Topic Area: Quality ",Draft 2 Encyclopedia of Software Engineering. See also members.aol.com/ONeillDon2/peer-reviews.html - 67k
[8] Bordin Sapsomboon ," Software Inspection and Computer Support", state of the art paper, 1999. www.sis.pitt.edu/~cascade/bordin/soa_inspection.pdf
[9] Ciolkowski, M. Laitenberger, O. Rombach, D. Shull, F. Perry, D. Fraunhofer, "Software inspections, reviews and walkthroughs",Proceedings of the 24th International Conference on Software Engineering, ICSE 2002. ISBN 1-58113-472-X, pg 641-642
[10] Lili Murtha, " Benefits of Software Inspections", Society for Software Quality 1999
[11] Software Metrics see sunset.usc.edu/classes/cs577b_2001/metricsguide/metrics.html - 25k
[12] Craig Borysowich , "Inspection/Review Meeting Metrics", 2006. see also blogs.ittoolbox.com/eai/implementation/archives/sample-inspectionreview-metrics-13640 - 184k
[13] Joe Schofield, "Beyond Defect Removal: Latent Defect Estimation with Capture Recapture Method (CRM)", published in IT Metrics and Productivity Journal, August 21, 2007
[14] Adam A. Porter, Carol A. Toman and Lawrence G Votta, " An

Experiment to Assess the Cost-Benefits of Code Inspections in Large Scale Software Development", IEEE Transactions on Software Engineering, Vol. 23, No. 6, June 1997

[15] Lars M. Karg, Arne Beckhaus, " Modelling Software Quality Costs by Adapting Established Methodologies of Mature Industries", Proceedings of 2007 IEEE International Conference in Industrial Engineering and Engineering Management in Singapore, ISBN 078-1-4244-1529-8, Pg 267-271, 2-4 Dec.2007

[16] K.S. Jasmine, R. Vasantha ,"DRE – A Quality Metric for Component based Software Products", proceedings of World Academy Of Science, Engineering And Technonolgy, Vol 23, ISSN 1307-6884, August 2007